\documentclass[aps,pra,twocolumn,amsmath,amssymb,showpacs,superscriptaddress,tightenlines]{revtex4}
\usepackage{graphicx}% Include figure files
\usepackage{epsfig}
\usepackage{dcolumn}% Align table columns on decimal point
\usepackage{bm}% bold math
\usepackage{amssymb}
\usepackage{bm}
\usepackage{amsmath}
\usepackage[colorlinks,citecolor=blue,linkcolor=blue,hyperindex,CJKbookmarks,dvipdfm]{hyperref}
\usepackage{CJK}

\begin{document}

\title{Phase-kicked control of counter-rotating interactions in the quantum Rabi model}

\author{Jin-Feng Huang}

\address{Department of Physics and Institute of Theoretical Physics, The Chinese
University of Hong Kong, Shatin, Hong Kong Special Administrative
Region, People's Republic of China}

\author{C. K. Law}

\address{Department of Physics and Institute of Theoretical Physics, The Chinese
University of Hong Kong, Shatin, Hong Kong Special Administrative
Region, People's Republic of China}

\date{\today}
\begin{abstract}
We present an interaction scheme to control counter-rotating terms in the quantum Rabi model. We show that by applying a sequence of $\pi/2$ phase kicks to a two-level atom and a single mode
quantized field, the natural dynamics of the Rabi model can be
interrupted in a way that counter-rotating
transitions can be significantly enhanced. This is achieved by a
suitable timing of the phase kicks determined by a phase matching
condition. If the time between successive kicks is sufficiently
short, our scheme is turned into a dynamical decoupling problem in
which the effects of counter-rotating terms can be strongly suppressed under ultrastrong coupling.
\end{abstract}


\maketitle

\section{Introduction}

In this paper we investigate a mechanism of controlling virtual
processes occurring in a two-level atom strongly interacting with
a single-mode quantized electromagnetic field. Virtual processes
here refer to excitation-number non-conserving processes due to
counter-rotating terms in the interaction Hamiltonian. Such
processes are generally fast oscillatory with feeble amplitudes
and therefore the employment of rotating wave approximation (RWA)
is justified when studying the quantum dynamics under the
framework of the Jaynes-Cummings (JC)
model~\cite{JC1963,Knight1993JMO}. In doing so, however,
interesting physics of virtual processes is lost. To go beyond the
JC model, recent investigations have begun to turn to the
ultrastrong coupling regime in which the vacuum Rabi frequency is
comparable to the atomic transition frequency and the field
frequency. In such a regime, RWA is no longer valid and countering
rotating terms can give rise to novel features such as quantum
integrability~\cite{Braak}, asymmetry of vacuum
Rabi-splitting~\cite{Cao2011NJP}, non-classical photon
statistics~\cite{Hartmann2012PRL,Hartmann2013PRL}, superradiance
transition~\cite{Ashhab2013PRA}, and the spontaneous release of
virtual photons~\cite{Savasta2013PRL,Huang2014PRA}. In addition,
counter-rotating terms can modify the collapse and revival
dynamics~\cite{Solano2010PRL}, and quantum Zeno and anti-Zeno
effects~\cite{Cao2010PRA,Ai2010PRA}.

We note that the realization of an ultrastrong coupling in natural
systems is still a challenge~\cite{Girvin}, although recent
experiments have demonstrated ultrastrong coupling in artificial
atomic systems with the vacuum Rabi frequency being a moderate
fraction of the field frequency~\cite{Mooij2010PRL,Gross2010NatPhy,Huber2009Nat,Beltram2009PRB,Sirtori2010PRL}.
Therefore an interesting question is how to enhance the
counter-rotating terms or the corresponding virtual transitions
without increasing the coupling strength to the ultrastrong
coupling regime. One possible solution is to use a fast modulation
of the coupling strength such that counter-rotating terms are
converted into co-rotating ones. This strategy has been considered
in Ref.~\cite{Liberato2009PRA} to generate quantum vacuum
radiation, and it requires the coupling strength to oscillate at a
frequency which is about twice the cavity field frequency.

In this paper we indicate an alternative approach based on a
sequence of phase kicks to the system. Each phase kick corresponds
to a unitary operation determined by the square root of the
parity operator. As we shall see below, by repeatedly applying
phase kicks to the system with a suitable time separation between
successive kicks, we can strongly enhance a certain virtual
transition for systems with an interaction strength far below the
ultrastrong coupling regime. We point out that the enhanced
virtual transition is induced by matching the phase of quantum
evolution rather than the traditional energy-level resonance, and
so a high frequency modulation of coupling strength is not
required.

Interestingly, if the time between adjacent kicks is sufficiently
short, counter-rotating terms can be significantly suppressed even
in the ultrastrong coupling regime. In other words, states of
different excitation numbers are dynamically decoupled. We note
that dynamical decoupling have been studied extensively in the
context of decoherence control~\cite{viola}. Our scheme here
generalizes the concept by Vitali and Tombesi~\cite{Tombesi1999PRA,Tombesi2001PRA} in order to suppress unwanted
transitions by counter-rotating terms. The problem is relevant to
systems in the ultrastrong coupling regime in which the Rabi
oscillations are distorted appreciably. With our scheme, we can
restore the JC dynamics effectively.

\section{The interaction scheme}

To begin with we consider a two-level atom interacting with a single-mode
cavity field. The system is modelled by the quantum Rabi model with
the Hamiltonian $(\hbar=1)$~\cite{Rabi}:
\begin{eqnarray}
H_{R} & = & H_{JC}+H_{V},\label{eq:HR-1}
\end{eqnarray}
where $H_{JC}$ is the JC Hamiltonian, and $H_{V}$ contains
counter-rotating terms:
\begin{eqnarray}
H_{JC} & = & \frac{\omega_{0}}{2}\sigma_{z}+\omega_{c}a^{\dagger}a+\lambda_{0}\left(a\sigma_{+}+a^{\dagger}\sigma_{-}\right),\\
H_{V} & = & \lambda_{0}(a^{\dagger}\sigma_{+}+a\sigma_{-}).
\end{eqnarray}
Here, $a$ and $a^{\dagger}$ are annihilation and creation operators
of the field mode, and the two-level atom has a ground state
$|g\rangle$ and an excited state $|e\rangle$ so that
$\sigma_{z}\equiv|e\rangle\langle e|-|g\rangle\langle g|$ is
defined as usual. The $\omega_{0}$ and $\omega_{c}$ are atomic and
cavity field frequencies respectively, and $\lambda_{0}$ denotes
the atom-cavity interaction strength which is half the vacuum
Rabi frequency. In this paper we shall focus on the near resonance
system in which $\omega_{c}\approx\omega_{0}$. The ground state of
$H_{JC}$ is denoted by $|g,0\rangle$ which refers to a
ground state atom in the vacuum field. For later purpose, we let
$\left|n,s\right\rangle $ be excited states of $H_{JC}$, i.e.,
$H_{JC}\left|n,s\right\rangle
=\epsilon_{n,s}\left|n,s\right\rangle $ ($n=1,2,3,\cdots,\:
s=\pm$) with $n$ being the excitation number (eigenvalues of
$a^{\dag}a+|e\rangle\langle e|$), and $s=+$ (-) labelling the
higher (lower) energy level of the doublet for a given $n$.

\begin{figure}
\includegraphics[bb=24bp 254bp 569bp 420bp,clip,scale=0.4]{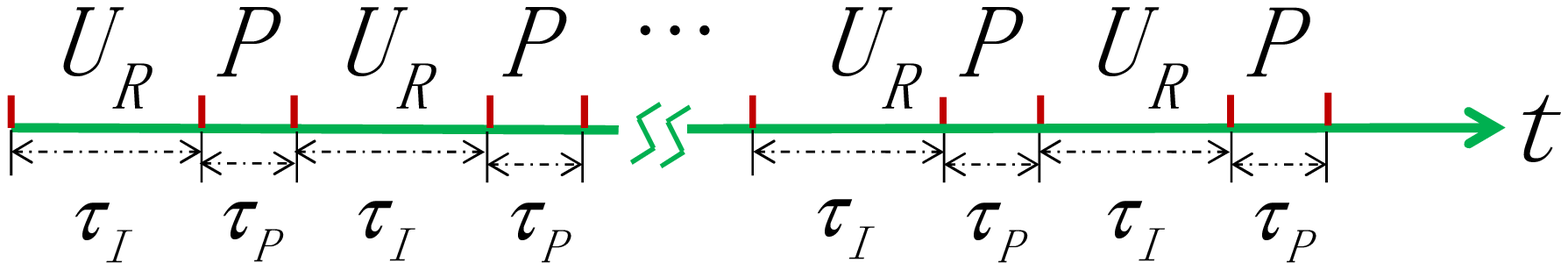}\caption{\label{fig:1}
Schematic diagram of the evolution sequence. $U_{R}=\textrm{exp}\left(-iH_{R}\tau_{I}\right)$
and $P$ denote the evolution governed by Rabi Hamiltonian and the
phase kick within time interval $\tau_{I}$ and $\tau_{P}$, respectively. }
\end{figure}

Our strategy to control the effects of $H_{V}$ is to introduce a
sequence of phase kicks to the system (Fig.~\ref{fig:1}). Each phase kick is
described by the unitary operator $P$:
\begin{equation}
P=\mbox{exp}\left[-i\pi(a^{\dagger}a+\sigma_{z}/2)/2\right],
\end{equation}
which is realized in a duration $\tau_{P}$. We assume that the system
evolution during each phase kick is entirely due to $P$. Therefore
the evolution operator of the system after $N$ phase kicks is given
by:
\begin{eqnarray}
U(NT) & = &
Pe^{-iH_{R}\tau_{I}}.....Pe^{-iH_{R}\tau_{I}}Pe^{-iH_{R}\tau_{I}}
\nonumber \\
&=& \left(Pe^{-iH_{R}\tau_{I}}\right)^{N},\label{eq:U1}
\end{eqnarray}
where $T=\tau_{I}+\tau_{P}$, and $\tau_{I}$ denotes duration of
evolution under $H_{R}$.

\section{Enhancement of counter-rotating transitions}

\begin{figure}
\includegraphics[bb=95bp 0bp 459bp 330bp,clip,scale=0.7]{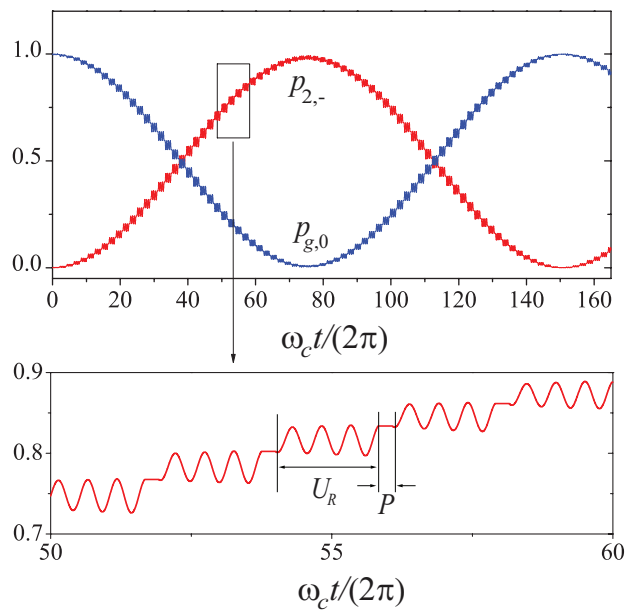}

\caption{ Occupation probabilities $p_{g,0}= |\langle g,0 | \psi
(t) \rangle|^2$ and $p_{2,-}= |\langle 2,- | \psi (t) \rangle|^2$
as a function of dimensionless time for the system under the
influence of phase kicks. The lower figure is a magnified plot of
$p_{2,-}$. The initial state is $|g,0 \rangle$. Parameters are
$\lambda_{0}=0.06\omega_{c}$, $\tau_{I}=7\pi/\Delta_{-}=11.4827/\omega_{c}$,
$\tau_{P}=\pi/(2\omega_{c})$, $\omega_{0}=\omega_{c}$.
}\label{fig:2}
\end{figure}

We consider the regime $\lambda_{0}/\omega_{c}\ll1$ in which the
ground state of $H_{R}$ is well approximated by $|g,0\rangle$
under RWA. However, by applying the sequence of phase kicks with
$\tau_{I}$ properly chosen, we find that a system initially
prepared in $|g,0\rangle$ is no longer stable but to evolve to
higher excited states of $H_{JC}$. Note that the $P$ operator
alone does not change the photon number and atomic excitations. An
example is given in Fig.~\ref{fig:2} in which the time-dependent
state $|\psi (t) \rangle = U(t)|g,0\rangle$ is calculated
numerically, and the figure shows the time-dependence of
occupation probabilities of $|g,0\rangle$ and $|2,-\rangle$. We
see that the system can make a transition to $|2,-\rangle$ almost
completely after some time, and this is impossible without applying the phase kicks.

In order to visualize the dynamics of the phase-kicked system, we
have plotted Fig.~\ref{fig:2} in continuous time (not only the
discrete time $NT$). A detailed picture is displayed in the lower
figure. We see that during each time interval when the system is
governed by $U_{R}$, the population of $|2,-\rangle$ is fast
oscillatory about a constant mean value. However, the action of
$P$ can interrupt the phase of the oscillations in the way that
the population of $|2,-\rangle$ climbs up after each kick. This feature continues until the probability of $p_{2,-}$ is almost one.

\begin{figure}
\includegraphics[bb=0bp 8bp 245bp 200bp,clip,scale=0.9]{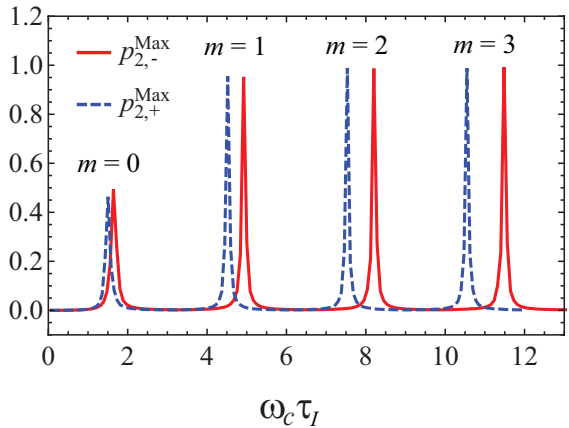}
\caption{\label{fig:3} The maximum probability
$p_{2,s}^{\textrm{Max}}$ ($s=\pm$) in the state
$\left|2,s\right\rangle $ as a function of $\omega_c \tau_{I}$.
Initial state is $\left|g,0\right\rangle $. Parameters are
$\lambda=0.06\omega_{c}$, $\omega_{0}=\omega_{c}$,
$\tau_{P}=\pi/(2\omega_{c})$.}
\end{figure}

The timings of phase kicks are important to obtain the enhanced
transition. In Fig.~\ref{fig:3}, we plot the maximum probability
$p_{2,\pm }^{\textrm{Max}}$ for the states $|2, \pm \rangle$
attainable as a function of $\tau_{I}$. The sharp peaks in
Fig.~\ref{fig:3} indicate that $p_{2,\pm }^{\textrm{Max}}$ are
significant only at certain values of $\tau_{I}$ located at
$\tau_{I}\approx(2m+1)\pi/\Delta_{\pm}$ ($m=0,1,2,\cdots$), where
$\Delta_{s}=\epsilon_{2,s}-\epsilon_{0}$ $(s=\pm)$ and
$\epsilon_{0}=-\omega_{0}/2$ is the ground state energy of
$H_{JC}$. These `resonance' values of $\tau_I$ correspond to a
phase matching condition that  $P$ is switched on at the moment
when $p_{2,s}$ reaches a turning (maximal) point, as illustrated
in the lower figure of Fig.~\ref{fig:2}.

Having presented our numerical observations, we now provide a
theory to explain the enhancement effect. Specifically we consider
the case of enhancing the transition from $|g,0\rangle$ to
$|2,s\rangle$ ($s=\pm$) with the required $\tau_{I}$ given by:
\begin{eqnarray}
\tau_{I}=[(2m+1)\pi+\phi]/\Delta_{s},\label{eq:En1}
\end{eqnarray}
where $m$ is a non-negative integer and $\phi\ll1$ is a small
constant angle. For definiteness, let us first examine the $s=-$
case. As Fig.~\ref{fig:2} suggests, the system is approximately
confined to the subspace formed by $|g,0\rangle$ and
$|2,-\rangle$, therefore we shall discuss the dynamics in this
subspace. The justification of such a two-level approximation will
be seen later.

The evolution operator for each cycle $T$ can be obtained approximately by using first-order time-dependent perturbation theory in which $H_{V}$ is treated as a perturbation in the interaction picture defined by $H_{JC}$. To the first order of $\lambda_{0}/\omega_{c}$, we have
\begin{equation}
Pe^{-iH_{R}\tau_{I}}\simeq e^{i\theta}(I-i{\cal
K}T),\label{eq:u-1}
\end{equation}
where $\theta=\pi/4-\epsilon_{0}\tau_{I}$ is a constant phase, and
${\cal K}$ is given by
\begin{eqnarray}
{\cal K}\simeq\frac{\phi}{T}\left|2,s\right\rangle \left\langle 2,s\right|
+\frac{2ig_{s}}{T}\left|2,s\right\rangle \left\langle g,0\right|-\frac{2ig_{s}}{T}\left|g,0\right\rangle \left\langle 2,s\right|\label{eq:Heff}
\end{eqnarray}
with $g_{s}=\lambda_{0}d_{2,s}/\Delta_{s}$, $d_{2,s}=\left\langle
e,1\right.\left|2,s\right\rangle $, and $I$ being the identity
operator. In deriving ${\cal K}$ in Eq. (\ref{eq:Heff}), we have made the
expansion exp$(i\phi)\approx1+i\phi$. After $N$ phase kicks, the
evolution operator in the two-level subspace is approximated by
\begin{eqnarray}
U\left(NT\right) & \approx & e^{iN \theta} e^{-i{\cal K}NT},\label{eq:U}
\end{eqnarray}
where the correction is $\mathcal{O}\left({N{\cal K}^{2}T^{2}}\right)$
which can be neglected if $N(g_{s}^{2}+\phi^{2})\ll1$.

Now we see that ${\cal K}$ in Eqs. (\ref{eq:Heff}-\ref{eq:U}) plays the role of an
effective two-level Hamiltonian governing the system evolution
after each kick. In particular, the terms $\phi/T$ and $g_{s}/T$
act as an effective detuning and Rabi frequency respectively. Therefore, on
one hand, $\phi=0$ corresponds to a resonance condition so that an
initial $|g,0\rangle$ state can be completely transferred to
$|2,-\rangle$ at the time $t=\pi T/\left(4g_{-1}\right)$, and this
agrees with the exact result obtained by numerical calculations in
Fig.~\ref{fig:2}; on the other hand, $\phi\gg g_{-}$ corresponds
to a large detuning situation in which the phase kicks have almost
no effects on the system. This explains the resonance curve in
Fig.~\ref{fig:3} because $\phi$ is related to $\tau_{I}$. In
particular, the sharp resonance peaks in Fig.~\ref{fig:3} are
located at $\tau_{I}$ corresponding to $m=0,1,2,...$ by Eq. (\ref{eq:En1}), and the peak width is $8\omega_{c}g_{s}/\Delta_{s}$.

The theoretical analysis above can be applied to the $|2,+\rangle$
transition by simply taking $s=+$ in Eqs. (\ref{eq:En1}-\ref{eq:U}), and this would
yield a different set of resonance. Note that in Fig.~\ref{fig:3},
the peaks associated with the $|2,+\rangle$ transition (blue
dashed curve) are well separated from those for the $|2,-\rangle$
transition (red curve), except for the $m=0$ case. To understand
this, let us choose $\tau_{I}$ at the exact resonance of the
$|2,-\rangle$ transition, i.e., $\tau_{I}=(2m+1)\pi/\Delta_{-}$,
then this value of $\tau_{I}$ is deviated from the resonance peak
of $|2,+\rangle$ of the same $m$ because
$\Delta_{+}\ne\Delta_{-}$. By Eq. (\ref{eq:En1}) with $s=+$, the deviation is
$\phi=(2m+1)\pi(\Delta_{+}-\Delta_{-})/\Delta_{-}$. Such a $\phi$
acts like an effective detuning which increases with $m$. In fact,
for the parameter used in Fig.~\ref{fig:3}, a large detuning condition $\phi
\gg g_{+}$ is satisfied when $m>1$, and hence the $|2,+\rangle$
transition is suppressed when $\tau_I$ is taken to be at the
resonance peak of the $|2,-\rangle$ transition. This explains why
peaks of different transitions in Fig.~\ref{fig:3} are more
separated for higher $m$'s, which also justifies our two-level
treatment in Eq. (\ref{eq:Heff}). The case of $m=0$ involves a partial overlap
of resonances, and it requires a multi-level treatment. However,
the details of this special case will not be discussed here.

\section{Suppression of counter-rotating transitions}

In this section we show that the sequence of phase kicks can be
used to suppress counter-rotating terms in the ultrastrong
coupling regime if $\tau_{I}<1/\omega_{0}$ is sufficiently short.
This is different from the previous section in which the values of
$\tau_{I}$ considered in Eq. (\ref{eq:En1}) can be much longer than
$1/\omega_{0}$ (or $1/\omega_{c}$). We begin by noting that the
parity operator of the Rabi model is given by
$\Pi=-\sigma_{z}(-1)^{a^{\dag}a}$ which commutes with $H_{R}$~\cite{Solano2010PRL}. By the fact that $\Pi=-iP^{2}$, we have
$P=P^{2}P^{\dag}=i\Pi P^{\dag}$, and so the evolution operator in
Eq. (\ref{eq:U1}) with an even $N$ is: $U(NT) =
(i\Pi)^{N/2}(P^{\dag}e^{-iH_{R}\tau_{I}}Pe^{-iH_{R}\tau_{I}})^{N/2}$,
since $[H_{R},\Pi]= [\Pi,P]=[\Pi,P^{\dag}]=0$. Next by
$P^{\dag}H_{R}P=H_{JC}-H_{V}$, we have
\begin{eqnarray}
U(NT)=(i\Pi)^{\frac{N}{2}}(e^{-i\left(H_{JC}-H_{V}\right)\tau_{I}}e^{-i\left(H_{JC}+H_{V}\right)\tau_{I}})^{\frac{N}{2}},
\end{eqnarray}
which indicates that the application of two successive phase kicks
can alternatively change the sign of $H_{V}$. If $\tau_{I}$ is sufficiently small, we can make the
following approximation up to the second order of $\tau_{I}$:
\begin{eqnarray}
e^{-i\left(H_{JC}-H_{V}\right)\tau_{I}}e^{-i\left(H_{JC}+H_{V}\right)\tau_{I}} & \simeq & e^{-i(H_{JC}+\hat{\varepsilon})2\tau_{I}},\label{eq:Us}
\end{eqnarray}
where $\hat{\varepsilon}=-i\tau_{I}[H_{JC},H_{V}]/2$. Therefore, in
the limit $\tau_{I}\to0$ (but keeping time $NT$ fixed), we have $\hat{\varepsilon}\rightarrow0$, and
$U(NT) \to (i\Pi)^{N/2}$exp$(-iH_{JC}N\tau_{I})$, i.e., $H_{V}$ is
completely suppressed. In this limit, the system is governed by an
effective JC Hamiltonian $H_{JC}'=\tau_{I}H_{JC}/T$ since $NT$ is
the total time.

\begin{figure}
\includegraphics[bb=0bp 5bp 408bp 331bp,clip,scale=0.55]{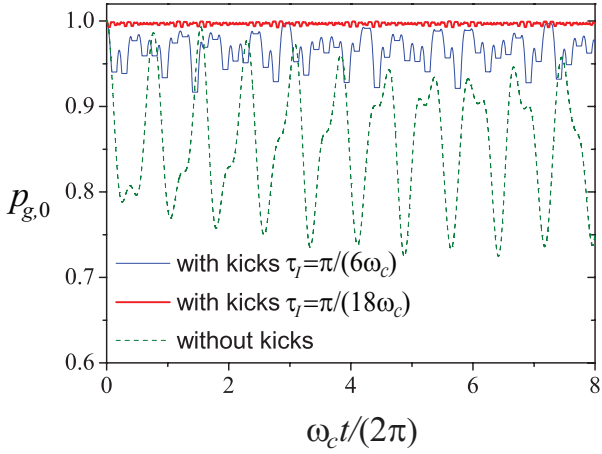}
\caption{Time-dependence of probabilities
$p_{g,0}$ for the cases: without phase kicks (green dashed curve),
with phase kicks using
$\tau_{I}=\tau_{P}=\pi/\left(6\omega_{c}\right)$ (blue thin solid
curve), and with phase kicks using
$\tau_{I}=\tau_{P}=\pi/\left(18\omega_{c}\right)$ (red thick solid
curve). The initial state is $\left|g,0\right\rangle $. Other
common parameters are $\omega_{0}=\omega_{c}$,
$\lambda_{0}=0.5\omega_{c}$. }\label{fig:pg0}
\end{figure}

To illustrate the suppression effect, we resort to numerical
calculations and show our results in Fig.~\ref{fig:pg0}. We
consider the system in the ultrastrong coupling regime so that
the effects of counter-rotating terms are prominent. With an initial
state given by $|g,0\rangle$, Fig.~\ref{fig:pg0} shows the
time-dependence of the probability of $|g,0 \rangle$  with and
without phase kicks. Because of the ultrastrong coupling,
$p_{g,0}$ for the case without kicks is oscillatory and it is
significantly deviated from one because of counter-rotating
transitions. However, by applying the phase-kick sequence, we see
that $p_{g,0}$ oscillation amplitudes are suppressed. In
particular, $p_{g,0} \approx 1$ for the case with a smaller
$\tau_I$ (red thick solid curve). This demonstrates that $|g,0 \rangle$ is
dynamically decoupled from the counter-rotating transitions.

\begin{figure}
\includegraphics[bb=0bp 12bp 251bp 359bp,clip,scale=0.8]{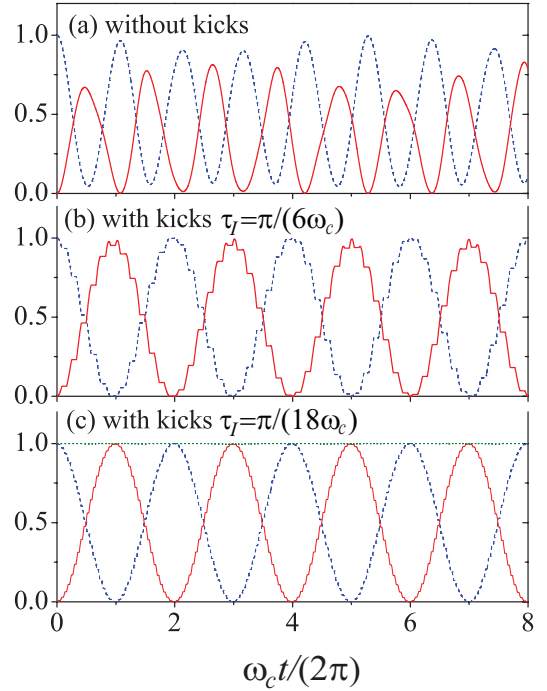}
\caption{\label{fig:4}Evolution of probabilities $p_{e,0}$ (blue
short-dashed curve) and $p_{g,1}$ (red solid curve) for states
$\left|e,0\right\rangle $ and $\left|g,1\right\rangle $,
respectively, with initial state $\left|e,0\right\rangle $.
Parameters are set as (a) without phase kicks; (b) with phase
kicks: $\tau_{I}=\tau_{P}=\pi/\left(6\omega_{c}\right)$; (c) with
phase kicks: $\tau_{I}=\tau_{P}=\pi/\left(18\omega_{c}\right)$.
The summation $p_{e,0}+p_{g,1}$ (green dot curve) is shown in (c).
Other common parameters are $\omega_{0}=\omega_{c}$,
$\lambda_{0}=0.5\omega_{c}$.}
\end{figure}

We also consider a different initial state given by $|e,0\rangle$,
Fig.~\ref{fig:4}(a) shows the evolution without phase kicks. We
see that the Rabi oscillations between $|e,0\rangle$ and
$|g,1\rangle$ are strongly modified by $H_{V}$.
In particular, the sum of probabilities of $|e,0\rangle$ and
$|g,1\rangle$ is not equal to one most of the time, which indicates
that virtual transitions beyond $|e,0\rangle$ and
$|g,1\rangle$ levels are significant. By using phase kicks {[}Fig.~\ref{fig:4}(b,c){]}, we see that sinusoidal Rabi oscillations are recovered.
These oscillations agree well with the prediction by the effective
JC Hamiltonian $H_{JC}'$ described above. In particular, a
smaller $\tau_{I}$ gives a better agreement {[}Fig.~\ref{fig:4}(c){]}. Note that the effective vacuum Rabi frequency is modified
to $\tau_{I}\lambda_{0}/T$.

We point out that our dynamical decoupling scheme is a
generalization of the parity-kick approach proposed by Vitali and
Tombesi~\cite{Tombesi1999PRA} in dealing with the decoherence-control
problems, and related ideas have been discussed in different
systems~\cite{Agarwal,Agarwal2011PRA}. In the case of our system,
we exploit the fact that the Rabi Hamiltonian has a parity symmetry,
i.e., $[H_{R},\Pi]=0$, and $P$ is effectively a parity kick on the
atom$+$field system. Such a kick can change the sign of the
coherence (off-diagonal elements of the density matrix)
associated with the unwanted virtual transition, and so reversing
the evolution due to $H_{V}$.

\begin{figure}
\includegraphics[bb=0bp 0bp 238bp 152bp,clip]{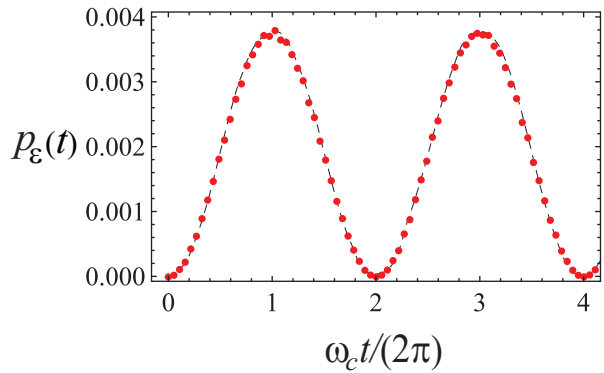}

\caption{\label{fig:6} The probability $p_{\varepsilon}(t)$
(defined in text) that measures the effect of $\hat{\varepsilon}$
operator. The black dashed curve is obtained by the first-order
time-dependent perturbation theory and red points are numerically
exact result. Same parameters as in Fig.~\ref{fig:4}(c).}
\end{figure}

Finally, we discuss the influence of  $\hat{\varepsilon}$ in Eq. (\ref{eq:Us}) when $\tau_{I}$ is finitely small.
Explicitly,
$\hat{\varepsilon}$ is given by,
\begin{eqnarray}
\hat{\varepsilon} & = & g_{1}a^{\dagger}\sigma_{+}+g_{2}a^{\dagger2}\sigma_{z}+\mbox{h.c.},
\end{eqnarray}
where $g_{1}=-i\tau_{I}\left(\omega_{0}+\omega_{c}\right)\lambda_{0}/2$
and $g_{2}=i\tau_{I}\lambda_{0}^{2}/2$ are defined. We can estimate
the correction due to $\hat{\varepsilon}$ by using first-order perturbation
theory, which gives the evolution operator
\begin{equation}
U(NT)\approx U_0
\left({1-i\int_{0}^{N\tau_{I}}{e^{iH_{JC}t'}}\hat{\varepsilon}e^{-iH_{JC}t'}dt'}\right),\label{eq:U2}
\end{equation}
where $U_0=(i\Pi)^{N/2}$exp$(-iH_{JC}N\tau_{I})$. By comparing
$H_{JC}$ and $\hat{\varepsilon}$, the correction term in the
bracket can be neglected as long as
$\left(\omega_{0}+\omega_{c}\right)\tau_{I}\ll1$ and
$\lambda_{0}\tau_{I}\ll1$. As an example, for the system evolving
from $|e,0\rangle$, $\hat{\varepsilon}$ can cause virtual
transitions to higher states $|3,\pm\rangle$. The probability of
excitation-number non-conserving transitions is characterized by
$p_{\varepsilon}(t)=1-|\langle e,0|\psi(t)\rangle|^{2}-|\langle
g,1|\psi(t)\rangle|^{2}$, where $|\psi(t)\rangle$ is the system
state. By Eq. (\ref{eq:U2}), we find that $p_{\varepsilon}$ is of order
$\tau_{I}^{2}\lambda_{0}^{2}$
 in the ultrastrong coupling regime. In Fig.~\ref{fig:6}, we plot $p_{\varepsilon}(t)$ by using
first-order perturbation theory Eq. (\ref{eq:U2}) and exact numerical
evolution operator in Eq. (\ref{eq:U1}), and they have a good agreement. For
the case $\tau_{I}=\pi/\left(18\omega_{c}\right)$ considered in
Fig.~\ref{fig:6}, $p_{\varepsilon}(t)$ is oscillatory with an
amplitude order of $10^{-3}$.

\section{Conclusion}

We have discovered a mechanism to control virtual transitions in
the quantum Rabi model. On one hand, virtual transitions can be
suppressed by using a sufficiently small $\tau_I$. Since the
dynamical decoupling only affects the counter-rotating terms, JC
dynamics can be recovered for the systems in the ultrastrong
coupling regime. On the other hand, virtual transitions can be
amplified by applying phase kicks at suitable timings.
Particularly, the enhanced virtual transitions can be exploited to
probe different energy levels by tuning $\tau_I$ as demonstrated
in Fig.~\ref{fig:3}. It is worth noting that the mechanism in Sec.
III can be employed as an excitation scheme for a general two-level
system in which $U_R$ and $P$ can be realized by non-resonant
classical pulses, and so a two-level system can be fully excited
without using a resonant driving field.

Our investigation suggests that the ability to imprinting quantum
phase changes with controllable timings could be a key to trigger
interesting quantum transitions. For our system, the operation
sequence in Eq. (\ref{eq:U1}) can be achieved by switching on and
off $\lambda_{0}$ with the duration $\tau_{I}$ and $\tau_{P}$
respectively. This is because when $\lambda_{0}=0$,
$H_{R}=\omega_{0}\sigma_{z}/2+\omega_{c}a^{\dagger}a$ is a free
Hamiltonian, and the corresponding free evolution operator can be
made equal to $P$ if $\tau_{P}$ is suitably chosen. For example,
in the resonance case of $\omega_{c}=\omega_{0}$, $P$ can be
achieved by switching off $\lambda_{0}$ in a duration
$\tau_{P}=(2m+1/2)\pi/\omega_{c}$, where $m$ is a non-negative
integer. We note that experimental progress of switchable coupling
for systems under ultrastrong coupling has been
made~\cite{Huber2009Nat,Ebbesen2011PRL,huber}, and there are also
theoretical studies on the designs and effects of switchable
coupling~\cite{GarciaRipoll2010PRL,Savasta2011PRL,Savasta2013PRA}.
We hope our work would motivate further investigations of the
subject in the future.

\begin{acknowledgments}
This work is partially supported by a grant from the Research Grants
Council of Hong Kong, Special Administrative Region of China (Project
No. CUHK401812).
\end{acknowledgments}

\end{document}